# Lagrangians for variational formulations of the Navier-Stokes equation

Sylvio R. Bistafa
sbistafa@usp.br
Retired, Polytechnic School, University of São Paulo
São Paulo, SP, Brazil

**Abstract**

Variational formulations for viscous flows which lead to the Navier-Stokes equation are examined. Since viscosity leads to dissipation and, therefore, to the irreversible transfer of mechanical energy to heat, thermal degrees of freedom have been included in the construction of viscous dissipative Lagrangians, by embedding of thermodynamics aspects of the flow, such as thermasy and flow exergy. Another approach is based on the presumption that the pressure gradient force is a constrained force, whose sole role is to maintain the continuity constraint, with a magnitude that is minimum at every instant. From these considerations, Lagrangians based on the minimal energy dissipation principal have been constructed from which the application of the Euler-Lagrange equation leads to the standard form of the Navier-Stokes equation directly, or at least they are capable of generating the same equations of motion for simple steady and unsteady 1 Dimensional viscous flows. These efforts show that there is equivalence between Lagrangian, Hamiltonian, and Newtonian mechanics as far as the derivation of the Navier-Stokes equation is concerned. However, one of the conclusions is that the attractiveness of the variational approach in more complex situations is still an open question for the applied fluid mechanician.

1. Introduction

The Navier–Stokes equation [1] arises from applying Newton's second law to fluid motion, together with the assumption that the stress in the fluid is the sum of a diffusing viscous term (proportional to the gradient of velocity) and a pressure term—hence describing viscous flow. The difference between the closely related Euler equation and the Navier–Stokes equation is that the latter takes viscosity into account while the former only model the inviscid flow.

Variational approaches to model viscous flows in fluid mechanics are based on the principle of minimum energy dissipation. This principle states that, for a given flow field, the rate of energy dissipation should be minimized. This can be achieved by finding the flow field that minimizes the dissipation rate, which is given by the rate of viscous dissipation in the fluid.

In order to do this, a functional is defined that describes the total energy dissipation in the fluid, and the flow field is found by minimizing this functional using techniques from the calculus of variations. This results in a set of partial differential equations that describe the flow. The dominant variational principle in physics is Hamilton's principle of least action, which does not directly allow for non-conservative forces (forces that do not come from a scalar work function). Therefore, most of the variational principles of fluid mechanics found in the literature have been developed for inviscid fluids governed by Euler's equations (e.g., [2, 3]), which do not take into consideration important features (e.g., viscosity, turbulence, and other irreversible phenomena). Nonetheless, the Hamiltonian structure has been mathematically manipulated to include artificial variables to recover the already known governing equations (e.g., [4, 5, 6]) but, without providing new insights on the physics of the fluid responsible for dissipation. Moreover, these derivations are often characterized by deep mathematical treatments, which tend to be rather detached from the physical aspects of the problem, and where unorthodox symbology and notation represents a barrier for the applied fluid mechanician. Also, it is often not clear at all how one can use any of these variational formulations to solve (even simple) viscous fluid mechanics problems.

Therefore, it is considered that a variational formulation for real flows must objectively include the dissipative nature of the viscous forces and its manifestations. To this end, as we shall see below, the formulations that have been proposed are in one form or another related to the principle that the viscous fluid motion has the minimal energy dissipation of any other motion consistent with the same boundary conditions — this is generally known as the Helmholtz minimum dissipation theorem.

In fluid mechanics, Helmholtz minimum dissipation theorem (named after Hermann von Helmholtz who published it in 1868) states that the steady Stokes flow motion of an incompressible fluid has the smallest rate of dissipation than any other incompressible motion with the same velocity on the boundary. The theorem also has been studied by Lord Rayleigh in 1913. This theorem is, in fact, true for any fluid motion where the nonlinear term of the incompressible Navier-Stokes equations can be neglected or equivalently when $\nabla \times \nabla \times \omega = 0$, where $\omega$ is the vorticity vector. For example, the theorem also applies to unidirectional flows such as Couette flow and the Hagen–Poiseuille flow, where the nonlinear terms disappear automatically. The Poiseuille flow theorem is a consequence of the Helmholtz theorem and states that *the steady laminar flow of an incompressible viscous fluid down a straight pipe of arbitrary cross-section is characterized by the property that its energy dissipation is least among all laminar (or spatially periodic) flows down the pipe which have the same total flux* (from the Wikipedia: https://en.wikipedia.org/wiki/Helmholtz_minimum_dissipation_theorem).

In the present paper, Lagrangian formulations for incompressible viscous flows that we have been able to identify from the point of view mentioned above are examined, to show the ingenuity of the proposals, to see how they are applied to solve classical viscous flow problems, and at the same time checking their attractiveness for further developments.

**2. The calculus of variations and the Euler-Lagrange equation**

The calculus of variations is a line of research that began with the proposal of the brachistochrone problem in 1690 by Jacob Bernoulli (1655–1705) who started a contest on finding the profile of a hanging flexible cord so that a bead will slide down the wire under gravity from one point to the other (without friction) in the shortest time. The resulting curve is known as the curve of fast descent or the brachistochrone curve.

This problem had been studied by several earlier investigators, including Leonhard Euler (1707–1783) who is considered the father of the variational calculus as laid out in his landmark 1744 book on variational techniques [7]. In this book, Euler also derives from geometrical arguments the so-called Euler-Lagrange equation in the form

$$\frac{\partial Z}{\partial y} - \frac{d}{dx}\left(\frac{\partial Z}{\partial y'}\right) = 0, \qquad (1)$$

from the requirement that the integral $\int Z dx$ must be a minimum, and where $Z$ is not a function, but instead a functional known as the *Lagrangian* $\mathcal{L}$. The solution of this differential equation gives the curve that minimizes the integral.

A particular example on the use of the Euler-Lagrange equation applied to the problem at hand may be taken from Seliger & Whitham [8], who considered that thermal degrees of freedom must be considered in a variational formulation of fluid flow which enters the formalism as the Lagrange multiplier $\vartheta$ corresponding to the entropy-conservation constraint; the proposed Lagrangian is then written as

$$\mathcal{L} = -\rho \left[ \frac{D\zeta}{Dt} + \alpha \frac{D\beta}{Dt} - s \frac{D\vartheta}{Dt} - \frac{\boldsymbol{u}^2}{2} + e(\rho, s) \right] \tag{2}$$

where

$$\boldsymbol{u} = \nabla \zeta + \propto \nabla \beta - s \nabla \vartheta, \tag{3}$$

which depends on the specific internal energy $e(\rho, s)$, given in terms of density $\rho$ and entropy $s$, the three of the so-called Clebsch potentials $\zeta, \propto, \beta$ [9] and an additional potential field $\vartheta$. For inviscid flows, the potential representation of the velocity field is given by $\boldsymbol{u} = \nabla \zeta + \propto \nabla \beta$.

To reveal the meaning of the thermal potential $\vartheta$ it is possible to apply the Euler–Lagrange equation with respect to variations in $s$, for which, $\frac{\partial Z}{\partial y} = \frac{\partial \mathcal{L}}{\partial s} = \frac{D\vartheta}{Dt} - \frac{\partial e}{\partial s} = 0$, giving

$$\frac{D\vartheta}{Dt} = \frac{\partial e}{\partial s} = T \tag{4}$$

or

$$\vartheta = \int_{t_0}^{t} T dt \tag{5}$$

where the integration is carried out along a particle trajectory. Because of its definition as given by Eq. (5), $\vartheta$ is often called the temperature integral, whereas Van Dantzig [10] called it the *thermasy*. From Eq. (5), It is seen that thermasy has units of temperature times time.

As pointed out by Scholle and Marner [11], although this approach is still restricted to adiabatic and therefore reversible processes, the above Lagrangian represents a momentous step forward because it is the first attempt of embedding of thermodynamics aspects of the flow. It is then seen that the most relevant difference to the classical potential theory is the appearance of an additional field, the thermasy $\vartheta$. According to these same authors, its physical meaning is not obvious: there are attempts in literature to relate this additional degree of freedom to a deviation from local thermodynamic equilibrium, however, there are still many open questions.

## 3. Construction of viscous dissipative Lagrangians

We examine next three attempts that have been proposed for the variational derivation of the Navier-Stokes equation from viscous dissipative Lagrangians. Two of these attempts may be called "thermal" approaches in which thermodynamic quantities are introduced into the Lagrangians to model the thermal dissipation of the viscous forces. Another attempt directly linked to the viscous force effect is based on the presumption that the pressure gradient force is a constrained force, which should be minimized at every instant.

**Thermasy-based extended Lagrangian**

Following the introduction of thermasy in the previous section, we shall first present Scholle and Marner's [11] extended Lagrangian proposal. Indeed, this has been accomplished by introducing terms modifying the entropy balance which results from the variation with respect to the thermasy $\vartheta$.

The production of entropy is considered through the dissipation of heat $\phi_d$ as $\vartheta\phi_d/T$, by assuming a quadratic dependence on $\frac{\partial u_i}{\partial x_j}$ according to the classical approach applied to viscous flows. The factor $1/T$ represents the character of the entropy as 'weighted heat' according to $\delta Q = Tds$. The extended Lagrangian in the absence of external forces and without considering the conduction of heat is then written as

$$\mathcal{L} = -\rho\left[\frac{D\zeta}{Dt} + \alpha\frac{D\beta}{Dt} - s\frac{D\vartheta}{Dt} - \frac{\boldsymbol{u}^2}{2} + e(\rho,s)\right] + \frac{\vartheta}{T}\left[\eta\,\mathrm{tr}\underline{D}^2 + \frac{\eta'}{2}(\nabla\cdot\boldsymbol{u})^2\right], \quad (6)$$

where $\eta$ is the shear viscosity, $\eta'$ is the volume viscosity of the fluid, and tr denotes the trace of the shear rate tensor $\underline{D}$.

From the variation with respect to the thermasy $\vartheta$, arises the entropy-balance equation given by

$$\frac{\partial(\rho s)}{\partial t} + \nabla\cdot(\rho s\boldsymbol{u}) = \frac{\eta}{T}\,\mathrm{tr}\underline{D}^2 + \frac{\eta'}{2T}(\nabla\cdot\boldsymbol{u})^2, \quad (7)$$

it is seen that the right-hand side of this equation represents the entropy production rate due to viscous dissipation, which implies that the second term in the right-hand side of Eq. (6) is the entropy production during the time interval $t - t_0$.

Momentum is of course a vector quantity whose density $p$ (momentum per unit volume) is $\rho\boldsymbol{u}$. Scholle and Marner's [11] revealed that the momentum density is now not simply equal to $\rho\boldsymbol{u}$, being the difference attributed to a quantity called the *quasi-momentum density* $p^*$, given by

$$p^* = -2\eta\nabla\cdot\left(\frac{\vartheta}{T}\underline{D}\right) - \eta'\nabla\left(\frac{\vartheta}{T}\nabla\cdot\boldsymbol{u}\right). \quad (8)$$

It is speculated that the quasi-momentum density could be due to the system's momentum balance beyond the scope of the continuum hypothesis on a molecular scale, e.g., Brownian motion.

The Euler-Lagrange equation is then applied by Scholle [12] to calculate the variations of the quantities that appear in the Lagrangian (Eq. 6), namely: $\boldsymbol{u},\vartheta,s,\rho,\alpha,\beta,\zeta$, which, after mathematical manipulations and other considerations, finally results in the equation of motion for the incompressible case given by

$$D_t\boldsymbol{u} = -\frac{\nabla p}{\rho} + \nu\{D_t + \nabla\otimes\boldsymbol{u}\}\left[2\underline{D}\nabla\left(\frac{\vartheta}{T}\right) + \frac{\vartheta}{T}\Delta\boldsymbol{u}\right] - \nu\,\mathrm{tr}\underline{D}^2\nabla\left(\frac{\vartheta}{T}\right), \quad (9)$$

where $D_t = \frac{\partial}{\partial t} + \boldsymbol{u}\cdot\nabla$, $\nu$ is the kinematic viscosity $\frac{\eta}{\rho}$, $\nabla\otimes\boldsymbol{u}$ is the velocity gradient tensor, and $\Delta\boldsymbol{u}$ is the Laplacian of the velocity vector $\boldsymbol{u}$.

It is seen that this equation of motion is not a reproduction of Navier-Stokes equation due to the occurrence of third-order derivatives instead of second-order terms, in a different form of the viscous terms, but also in an additional field, the thermasy $\vartheta$, appearing explicitly. The additional terms and degrees of freedom in Eq. (9) may represent an extension of the classical theory towards non-equilibrium thermodynamics.

Three benchmark tests of Eq. (9) have been performed by Scholle and Marner [11], which were earlier discussed in more details by Scholle [12], against exact solutions of the Navier-Stokes equation: (i) Stoke's first problem, (ii) plane Couette flow, and (iii) the Lamb-Oseen vortex diffusion. For these flows, a particular form of Eq. (9) was used

$$D_t \boldsymbol{u} = -\frac{\nabla p}{\rho} + \nu \{D_t + \nabla \otimes \boldsymbol{u}\} \left[\frac{\vartheta}{T} \Delta \boldsymbol{u}\right]. \tag{10}$$

which considers a particular solution of Eq. (5) given by $\frac{\vartheta}{T} = t - t_0$, and then, $D_t\left(\frac{\vartheta}{T}\right) = 1$.

The Stoke's first problem is a transient flow for which $\boldsymbol{u} = u(y, t)$, and by considering that $\nabla p = 0$, the $x$-component of Eq. (10) reads (noting also that $\nabla \otimes \boldsymbol{u}$ reduces to $\frac{\partial u}{\partial x} = 0$)

$$\frac{\partial u}{\partial t} = \nu \frac{\partial}{\partial t}\left[(t - t_0)\frac{\partial^2 u}{\partial y^2}\right]. \tag{11}$$

This third-order partial differential equation allows for integration with respect to time, leading to the equation

$$\frac{(u - u_0)}{(t - t_0)} = \nu \frac{\partial^2 u}{\partial y^2}{}^{\text{a}}, \tag{12}$$

whereas the Navier-Stokes equation leads to

$$\frac{\partial u}{\partial t} = \nu \frac{\partial^2 u}{\partial y^2}. \tag{13}$$

Although both equations are different, they imply a qualitatively similar evolution of velocity profiles. Eq. (13) is a classical diffusion equation whereas Eq. (12) is obtained by replacing the time derivative by a finite difference [13].

In the case of the plane Couette flow, we have additionally that the steady state solution is obtained in the limit $t \rightarrow \infty$, which transforms Eq. (12) into

$$0 = \frac{\partial^2 u}{\partial y^2}. \tag{14}$$

This is the same equation that would have been obtained by the direct application of the Navier-Stokes equation to the plane Couette flow [13].

Another example of transient flow is the Lamb-Oseen vortex diffusion where the velocity in cylindrical coordinates is $\boldsymbol{u} = u(r, t)$, and for which the vorticity $\omega$ is given by $\omega = \frac{1}{r}\frac{\partial (ru)}{\partial r}$; then from Eq. (11) we have that

$$\frac{\partial u}{\partial t} = \nu \frac{\partial}{\partial t}\left[(t - t_0)\frac{\partial \omega}{\partial r}\right]. \tag{15}$$

The integration of this equation with respect to time leads to

$$\frac{(u - u_0)}{(t - t_0)} = \nu \frac{\partial \omega}{\partial r}. \tag{16}$$

If at the initial time at the initial time $t_0 = 0$ and $u_0 = \frac{\Gamma}{2\pi r}$ where $\Gamma$ denotes its circulation, then, by applying the operator $\frac{1}{r}\frac{\partial (r \cdot)}{\partial r}$ to Eq. (16), we will have that

$$\frac{\omega}{t} = \nu \left( \frac{\partial^2 \omega}{\partial r^2} + \frac{1}{r} \frac{\partial \omega}{\partial r} \right). \tag{17}$$

---

[a] In fact, this equation was originally written as $\frac{u-u_0}{t-t_0} = f\nu \frac{\partial^2 u}{\partial y^2}$, where the factor $f$ was not clearly identified by Scholle [12]. This factor introduces a slight variation in the velocity profile characterized by a weaker spatial decay of the flow velocity when compared to that of the solution of the original Navier-Stokes equation. In the tests performed, the factor $f$ was found to be equal to $\pi$.

In this equation, $\frac{\omega}{t}$ has been replaced for $\frac{\partial \omega}{\partial t}$ which would have been obtained by the direct application of the Navier-Stokes equation, resulting in the well-known equation for the diffusion of a vortex [14].

Scholle [12] then showed that there is a qualitative agreement between the solution given by Eq. (17), and the solution of the original vortex diffusion equation as given by the Navier-Stokes equation, despite quantitative differences in the respective flow profiles.

The main finding, however, is that by the Lagrangian (Eq. 6) is able to capture the main features of these flow-tests.

Additional tests were performed by Scholle and Marner [11] for pressure driven flows for which no adequate solutions of the field equations could be constructed which simultaneously fulfil the pressure boundary conditions. The conclusion was that the variational principle based on the proposed Lagrangian (Eq. 6) does not recover the dynamics of viscous flow in a proper way, since its applicability seems to be restricted to special flow problems only.

From these examples another issue arose, since the explicit appearance of the weighted thermasy $\frac{\vartheta}{T}$ turns out to be of unlimited growth, either spatially or temporally, which also prohibits its interpretation in connection with non-equilibrium thermodynamics. Next, to overcome the above-highlighted anomalies, a modification of the proposed Lagrangian has been provided Scholle and Marner's [11], which is not considered further here.

**Flow exergy-based Lagrangian**

Sciubba [15] proposed a Lagrangian based on the exergy variation of the flow, to obtain the Navier-Stokes equation from the minimization of the exergy destruction. Exergy is a thermodynamic concept, used for many years within engineering analyses of chemical and mechanical processes and systems. Exergy is defined as *the maximum useful work which can be extracted from a system as it reversibly comes into equilibrium with its environment.* In other words, it is the capacity of energy to do the actual physical work.

The flow of a viscous fluid is driven by a set of well-defined external fields (pressure, external force, and temperature) and by the inertia of the mass under consideration, and it is affected by dissipative effects related to the viscosity and thermal conductivity of the fluid. Dissipation is associated with entropy production or to the exergy destruction of the flow. Flow exergy $e$ is an extensive thermodynamic state function defined as

$$e = h - h_0 - T_0(s - s_0), \tag{18}$$

where $h$ is the enthalpy, $s$ is the entropy and $T_0$ is a reference state temperature.

A representation of the work and heat interactions of a system in terms of exergy has the advantage of unifying both work/heat interactions and dissipative effects into a unified framework. Thus, for any dissipative system, a theorem of "exergy destruction" applies, which states that if the system undergoes an irreversible process, its specific exergy content is destroyed (annihilated) at a rate given by

$$\dot{e}_\lambda = T_0 \dot{s}_{irr}, \tag{19}$$

which states that every real (irreversible) process destroys exergy at a rate proportional to the irreversible entropy generation $\dot{s}_{irr}$.

In an element of time $dt$ the exergetic content per unit of mass is modified by four different contributions:

- an exergy variation rate equal to the reversible exchanged mechanical power $\dot{e}_{W_{rev}}$

$$\dot{e}_{W_{rev}} = \boldsymbol{u} \cdot \frac{D\boldsymbol{u}}{Dt} + \boldsymbol{u} \cdot \frac{\nabla p}{\rho} + \boldsymbol{u} \cdot \boldsymbol{B}, \tag{20}$$

where $\boldsymbol{B}$ is the body force.

- an exergy destruction rate proportional to the viscous dissipation function $D_{visc}$

$$\dot{e}_{\lambda_{visc}} = -\nu D_{visc}, \tag{21}$$

- an exergy variation rate proportional to the reversible thermal entropy exchange $\dot{s}_{rev}$

$$\dot{e}_{Q_{rev}} = (T - T_0)\dot{s}_{rev}, \tag{22}$$

- and an exergy destruction rate proportional to the irreversible thermal entropy production $\dot{s}_{irr,therm}$

$$\dot{e}_{\lambda_{therm}} = -(T - T_0)\dot{s}_{irr,therm}. \tag{23}$$

Thus, the total exergy variation per unit mass of the fluid $\Delta_{e_{fluid}}$ in time $dt$ is

$$\Delta_{e_{fluid}} = dt \left[ \boldsymbol{u} \cdot \frac{D\boldsymbol{u}}{Dt} + \boldsymbol{u} \cdot \frac{\nabla p}{\rho} + \boldsymbol{u} \cdot \boldsymbol{B} - \nu D_{visc} + (T - T_0)\dot{s}_{rev} - (T - T_0)\dot{s}_{irr,therm} \right]. \tag{24}$$

Once the flow variables are exactly known at each instant of time $t$ and at each point in the flow domain, the quantity defined by Eq (24) can be computed exactly locally and, if necessary, integrated over the entire domain to yield the global variation of the exergy of the flow.

Sciubba [15] then considers that if at every instant of time, the fluid motion is governed by the minimization of the exergy destruction given by Eq. (24), the resulting equation of motion is indeed the Navier-Stokes equation.

From Eq. (24), for an isothermal flow of a viscous homogeneous fluid with constant properties the Lagrangian can be written as

$$\mathcal{L} = \boldsymbol{u} \frac{D\boldsymbol{u}}{Dt} + \boldsymbol{u} \cdot \frac{\nabla p}{\rho} + \boldsymbol{u} \cdot \boldsymbol{B} - \nu D_{visc}, \tag{25}$$

which in index notation reads

$$\mathcal{L} = u_j \frac{\partial u_j}{\partial t} + u_j u_k \frac{\partial u_j}{\partial x_k} + \frac{1}{\rho} u_j \frac{\partial p}{\partial x_j} + u_j B_j - \nu D_{visc}, \tag{26}$$

where, as usual,

$$D_{visc} = \frac{1}{2} \nu \left( \frac{\partial u_i}{\partial x_j} + \frac{\partial u_j}{\partial x_i} \right)^2. \tag{27}$$

Imposing the condition of constrained minimum exergy destruction is equivalent to searching for the minimization of a functional whose integrand is the total exergy variation of the unit fluid mass given by Eq. (24); that is, $\int \mathcal{L} dV$ must be a minimum in $V$.

From Eq. (1), the Euler-Lagrange equation in index notation for the problem so posed reads

$$\frac{\partial \mathcal{L}}{\partial u_j} - \frac{\partial}{\partial x_i}\left[\frac{\partial \mathcal{L}}{\partial(\partial u_j/\partial x_i)}\right] = 0. \tag{28}$$

where

$$\frac{\partial \mathcal{L}}{\partial u_j} = \frac{\partial u_j}{\partial t} + u_k\frac{\partial u_j}{\partial x_k} + \frac{1}{\rho}\frac{\partial p}{\partial x_j} + B_j. \tag{29}$$

$$\frac{\partial \mathcal{L}}{\partial\left(\frac{\partial u_j}{\partial x_i}\right)} = \frac{\partial D_{visc}}{\partial\left(\frac{\partial u_j}{\partial x_i}\right)} = \nu\left(\frac{\partial u_i}{\partial x_j} + \frac{\partial u_j}{\partial x_i}\right), \tag{30}$$

from which

$$\frac{\partial}{\partial x_i}\left[\frac{\partial \mathcal{L}}{\partial\left(\frac{\partial u_j}{\partial x_i}\right)}\right] = \frac{\partial}{\partial x_i}\left[\nu\left(\frac{\partial u_i}{\partial x_j} + \frac{\partial u_j}{\partial x_i}\right)\right] = \nu\left[\frac{\partial}{\partial x_j}\left(\frac{\partial u_i}{\partial x_i}\right) + \frac{\partial^2 u_j}{\partial x_i \partial x_i}\right] = \nu\frac{\partial^2 u_j}{\partial x_i \partial x_i}, \tag{31}$$

since for incompressible flows $\left(\frac{\partial u_i}{\partial x_i}\right) = 0$.

A direct substitution of Eqs. (29) and (31) into Eq. (28) results in

$$\frac{\partial u_j}{\partial t} + u_k\frac{\partial u_j}{\partial x_k} + \frac{1}{\rho}\frac{\partial p}{\partial x_j} + B_j - \nu\left(\frac{\partial^2 u_j}{\partial x_i \partial x_i}\right) = 0, \tag{32}$$

which is indeed the Navier-Stokes equation for an incompressible isothermal flow.

The method is considered "restricted" by Sciubba [15], in the sense that the minimization was performed in space, but not in time, i.e., the time derivative of the velocity is not subjected to the variation which corresponds to the steady flow.

**Minimum pressure gradient-based Lagrangian**

Taha and Gonzalez [16] have sought to transform any fluid mechanics problem into an optimization one with no need to apply the Navier-Stokes equation. With this goal in mind, they have applied Gauss's principle of least constraint[b] to find the equation of motion of incompressible viscous flows. This principle is similar to Hamilton's principle which states that the true path taken by a mechanical system is an extremum of the action.

The principle of least constraint is a least squares principle stating that the true accelerations of a mechanical system of $n$ masses is the minimum of the quantity $Z$ given by

$$Z = \sum_{i=1}^{n} m_i\left(\boldsymbol{a}_i - \frac{\boldsymbol{F}_i}{m_i}\right)^2, \tag{33}$$

where $m_i$ is the mass of the *i*th particle, $\boldsymbol{a}_i$ is the corresponding acceleration which satisfy the imposed constraints, and which in general depends on the current state of the system, and $\boldsymbol{F}_i$ is the non-constraint force applied to the *i*th particle.

---

[b] The principle of least constraint is one variational formulation of classical mechanics enunciated by Carl Friedrich Gauss in 1829, equivalent to all other formulations of analytical mechanics. Intuitively, it says that the acceleration of a constrained physical system will be as similar as possible to that of the corresponding unconstrained system (from the Wikipedia: https://en.wikipedia.org/wiki/Gauss%27s_principle_of_least_constraint).

These authors showed that the pressure gradient force is a constrained force, whose sole role is to maintain the continuity constraint, with a magnitude that is minimum at every instant. Then, by considering that the pressure gradient is a constraint force, and the viscous force is an impressed force, the action $\mathcal{A}$ (Gauss' $Z$ quantity) was written as

$$\mathcal{A} = \frac{1}{2}\int \rho \left(\frac{D\boldsymbol{u}}{Dt} - \nu\Delta\boldsymbol{u}\right)^2 d\boldsymbol{x}, \qquad (34)$$

where $\frac{D\boldsymbol{u}}{Dt} = \frac{\partial \boldsymbol{u}}{\partial t} + \boldsymbol{u}\cdot\nabla\boldsymbol{u}$.

These authors pointed out that "… $\mathcal{A}$ is simply equal to $\int(\nabla p)^2 d\boldsymbol{x}$, and since the pressure force is a constraint force (enforcing the continuity constraint), the flow field will deviate from the motion dictated by the inertia $\boldsymbol{u}\cdot\nabla\boldsymbol{u}$ and viscous $\nu\Delta\boldsymbol{u}$ forces only by the amount to satisfy continuity; no larger pressure gradient will be generated than that necessary to maintain continuity. Nature will not overdo it. This new principle is what we call *The Principle of Minimum Pressure Gradient* (PMPG)." Therefore, it is seen that the action $\mathcal{A}$ seeks automatically the minimization of the pressure gradient.

The variational principle of minimum pressure gradient (PMPG) was then applied to solve two classical viscous flow problems in fluid mechanics — performing pure optimization without resorting to Navier-Stokes equation, namely: channel flow, and Stoke's second problem —, by Taha and Gonzalez [16]. The Euler-Lagrange equation (Eq. 1) comes from the requirement that $\int Z d\boldsymbol{x}$ must be a minimum, where from Eq. (33)

$$Z = \frac{1}{2}\rho\left(\frac{D\boldsymbol{u}}{Dt} - \nu\Delta\boldsymbol{u}\right)^2. \qquad (35)$$

For the channel flow, $u = u(y)$, the functional $Z$ given by Eq. (35) reduces to the Lagrangian $\mathcal{L} = \frac{1}{2}\left(\mu\frac{\partial^2 u}{\partial y^2}\right)^2$. Since this Lagrangian contains second-order derivatives, then the *Euler-Poisson* equation applies, which here is written as

$$\frac{\partial \mathcal{L}}{\partial u} - \frac{\partial}{\partial y}\left[\frac{\partial \mathcal{L}}{\partial\left(\frac{\partial u}{\partial y}\right)}\right] + \frac{\partial^2}{\partial y^2}\left[\frac{\partial \mathcal{L}}{\partial\left(\frac{\partial^2 u}{\partial y^2}\right)}\right] = 0 \qquad (36)$$

giving

$$\mu\frac{\partial^2 u}{\partial y^2} = constant. \qquad (37)$$

Also, according to PMPG, we also have that

$$\int (\nabla p)^2 d\boldsymbol{x} = 0, \qquad (38)$$

which from Euler-Lagrange equation results in

$$\frac{\partial}{\partial y}\left[\left(\frac{\partial p}{\partial x}\right)^2\right] = 0 \Longrightarrow \frac{\partial p}{\partial x} = constant. \qquad (39)$$

The equality between Eqs. (37) and (39) then gives

$$\mu\frac{\partial^2 u}{\partial y^2} = \frac{\partial p}{\partial x}, \qquad (40)$$

which is the same equation that is obtained by the direct application of the Navier-Stokes equation to the channel flow of the incompressible viscous fluid [13].

The Stokes' second problem is the unsteady flow above a harmonically-oscillating, infinitely-long plate, where $u = u(y, t)$, for which the functional $Z$ given by Eq. (35) reduces to the Lagrangian $\mathcal{L} = \frac{1}{2}\rho \left(\frac{\partial u}{\partial t} - \nu \frac{\partial^2 u}{\partial y^2}\right)^2$. This is an unsteady problem, for which the variation in the PMPG is taken with respect to the local acceleration, where the Euler-Lagrange equation is written as

$$\frac{\partial}{\partial t}\left[\frac{\partial \mathcal{L}}{\partial \left(\frac{\partial u}{\partial t}\right)}\right] = 0 \tag{41}$$

giving

$$\frac{\partial u}{\partial t} - \nu \frac{\partial^2 u}{\partial y^2} = constant. \tag{42}$$

From Eq. (39) above, $\frac{\partial p}{\partial x} = constant$, where now the constant should be equal to zero because there is no pressure gradient in the $x$-direction for this flow, therefore,

$$\frac{\partial u}{\partial t} = \nu \frac{\partial^2 u}{\partial y^2}, \tag{43}$$

which is recognized as the same equation that is obtained by the direct application of the Navier-Stokes equation [13]. It should be noted that Eq. (43) is equally applicable to the Stoke's first problem, which is another type of unsteady flow in which the plate is suddenly jerked into motion in its own plane with a constant velocity.

Although not within the context of viscous flows, and to show the attractiveness of variational formulations, Taha and Gonzalez [16] also addressed the so-called 'airfoil problem' showing that the inviscid version of PMPG is also capable of providing a unique solution for the flow over arbitrarily smooth shapes. As is well known, the potential flow over a two-dimensional object does not have a unique solution. As they put it, the only theoretical fix available is the so-called *Kutta condition*, which dictates the amount of circulation necessary to remove the singularity such as a sharp trailing edge in a conventional airfoil. However, for singularity free shapes (e.g., ellipse, circle), the Kutta condition is not applicable; and there is no theoretical model that can predict circulation and lift over these shapes. To circumvent this problem, these authors have proposed a novel variational formulation to Euler's equation for the dynamics of inviscid fluids where the *Appellian* (integral of squared acceleration) is minimized.

For the inviscid incompressible steady flow, the Appellian $\mathcal{A}$ may be written as

$$\mathcal{A}(\Gamma) = \frac{1}{2}\rho \int [\boldsymbol{u}(\boldsymbol{x}, \Gamma) \cdot \boldsymbol{\nabla} \boldsymbol{u}(\boldsymbol{x}, \Gamma)]^2 d\boldsymbol{x}. \tag{44}$$

Since the velocity field $\boldsymbol{u}(\boldsymbol{x}, \Gamma)$ is known except for the unknown parameter $\Gamma$, given an assumed value of $\Gamma$, one can compute the flow field and, consequently, the scalar Appellian integral $\mathcal{A}$ in Eq. (44).

The ArgMin of Eq. (44) is then used to find the smallest possible value $\Gamma^*$ of the circulation for the given constraints, providing a generalization of the Kutta-Zhukovsky condition that is, unlike the latter, derived from first principles. The PMPG allows, for the first time, computation of lift over smooth shapes without sharp edges where the Kutta condition fails.

Next, a family of airfoils parameterized by $D$ have been considered Taha and Gonzalez [17], in which $D$ controls smoothness of the trailing edge: $D = 0$ results in the classical Zhukovsky airfoil with a sharp trailing edge, and $D = 1$ results in a circular cylinder. From an involving calculation procedure, a plot of the locus of the minimizing circulation $\Gamma^*$ was eventually generated showing that for $D = 0$ (sharp-edged airfoil), the minimizing circulation $\Gamma^*$ coincides with Kutta's circulation; and for $D = 1$ (circular cylinder), the minimizing circulation vanishes, implying no inviscid lifting capability of this purely symmetric shape.

Finally, in a recent publication, Taha and Gonzalez [18] claim to have provided a theorem that establishes connection between the PMPG and Navier-Stokes equation in the general case, not only for specific examples, which is not considered further here.

## 4. Summary and Discussion

Three attempts to construct viscous dissipative Lagrangians to recover the Navier-Stokes equation from variational principles have been examined. In one form or another, these attempts are all related to the principle that the viscous fluid motion has the minimal energy dissipation of any other motion consistent with the same boundary conditions, which is a principle that appears in the literature under different statements: Helmholtz minimum dissipation theorem, Poiseuille flow theorem, and Gauss's principle of least constraint.

Since viscosity leads to dissipation and therefore to the irreversible transfer of mechanical energy to heat, thermal degrees of freedom have been included in the construction of viscous dissipative Lagrangians. One of these attempts adds to the potential representation of the Lagrangian a rather unknown thermodynamic quantity called thermasy (or temperature integral), whose physical meaning seems to be related to a deviation from local thermodynamic equilibrium. It is shown that the variation with respect to the thermasy results in an entropy balance equation where the entropy production rate is due to dissipation, as expected. Three benchmark tests have been performed with this extended Lagrangian for simple steady and unsteady 1 Dimensional flows, showing that it could lead to differential equations with the same character to those obtained from the direct application of the Navier-Stokes equation. However, additional tests performed by the authors led to the conclusion that the variational principle based on the proposed Lagrangian is not a general one, since it does not recover the dynamics of other viscous flows tested in a proper way, and, therefore, its applicability seems to be restricted to special flow problems only.

In another "thermal" approach, the Lagrangian density is obtained from the exergy balance equation written for the isothermal incompressible flow. The exergy of a fluid mass, composed of a kinetic, pressure work and body force work, reversible and irreversible thermal entropy exchange, and a dissipative portion, the latter being the result of viscous irreversibility is derived first, and it is then shown that a formal minimization of the exergy variation (i.e., destruction) generates a Lagrangian which, by applying the Euler-Lagrange equation to it, leads directly to the standard form of the Navier-Stokes equation for the isothermal incompressible flow of a viscous homogeneous fluid. However, the author of this approach considers that new applications will have to be developed for specific cases and tested experimentally, analytically, and numerically in unknown flow fields, and that an even more important step would be that of extending its validity to nonisothermal and compressible flows.

Finally, an approach was examined which is based on the presumption that the pressure gradient force is a constrained force due to the viscous force, whose sole role is to maintain the continuity constraint, with a magnitude that is minimum at every instant (Principle of Minimum Pressure Gradient, PMPG). A Lagrangian was then proposed, considering that the pressure gradient is a constraint force, and that the viscous force is an impressed force. From the equality between the Lagrangian expressing the minimal of the pressure gradient, it is then shown by the application of the Euler-Lagrange equation that the same differential equations are generated as those obtained from the direct application of the Navier-Stokes equation to simple steady and unsteady 1 Dimensional flows. The authors consider that the PMPG could shed light on the problem of existence of solutions of Navier-Stokes equation, because variational principles have usually been useful in studying existence of solutions of partial differential equations.

5. **Conclusions**

The proposals examined here show that a description of viscous flows is possible within the framework of the Lagrangian formalism, which leads to the standard form of the Navier-Stokes equation directly, or at least it is capable of generating the same equations of motion for simple steady and unsteady 1 Dimensional flows.

Perhaps, the most important revelation of the present paper is to show the power of the Euler-Lagrange equation in generating the Navier-Stokes equation once the key physical phenomenon involved (here viscous dissipation and its manifestations) has been properly modeled and included in the Lagrangians. However, these may lead to third-order derivatives instead of second-order terms, which have posed no additional difficulties for the simple flow problems considered here, but that may not be so in more complex situations.

As far as the derivation of Navier-Stokes equation is concerned, it is shown that there is equivalence between Lagrangian, Hamiltonian, and Newtonian mechanics, which, however, does not imply that every formulation of a physical problem is equally tractable in each framework.

Although the variational approaches examined here have proven to be capable of solving simple flow problems, their attractiveness in more complex situations is still an open question for the applied fluid mechanician.